\documentclass[%
superscriptaddress,
preprint,
 amsmath,amssymb,
 aps,prl,
]{revtex4-2}

\usepackage{graphicx}% Include figure files
\usepackage{dcolumn}% Align table columns on decimal point
\usepackage{bm}% bold math
\usepackage{xcolor}
\begin{document}

%\preprint{APS/123-QED}

\title{Mass-conserving growth percolation in polymer gelation}% Force line breaks with \\

\author{Ameya Rege}
\email{ameya.rege@utwente.nl}
\affiliation{%
 University of Twente, P.O. Box 217, Enschede, 7500 AE, The Netherlands
}%
\affiliation{%
 German Aerospace Center (DLR), Linder Höhe, 51147, Cologne, Germany
}%

\author{Lorenz Ratke}%
\affiliation{%
 German Aerospace Center (DLR), Linder Höhe, 51147, Cologne, Germany
}%

\date{\today}% It is always \today, today,

\begin{abstract}
Polymer gelation involves the emergence of a system-spanning network from growing polymer-rich domains, yet conventional percolation models typically prescribe particle size independently of material consumption. We formulate gelation as a locally mass-conserving growth-percolation process in which Voronoi capture zones define finite material reservoirs for individual nuclei. Local depletion determines the evolving supersaturation and limiting particle size, while particle contacts generate a dynamic network whose first spanning cluster defines the gel point. Three mechanisms, namely, interface-, diffusion-, and polymer-blob-controlled growth produce distinct gelation kinetics while sharing the same accessible final state. Spatial fluctuations in nucleation further generate a distribution of gelation times. The framework separates gelation from saturation and naturally captures post-gel aging as continued growth and topological maturation.
\end{abstract}

\maketitle

%\tableofcontents

Polymer gels, aerogels, and colloidal networks, form through the progressive growth and interconnection of initially isolated polymer-rich domains \cite{ratke2021chemistry}. Stable nuclei consume monomers from the surrounding solution, grow in size, establish contacts with neighboring domains, and eventually form a system-spanning network. The appearance of this spanning structure marks the transition from a dispersed collection of particles to a connected solid skeleton and may therefore be interpreted as a growth-driven percolation process. Predicting this transition requires a description that connects local particle growth and material consumption to the global topology of the emerging network.

Classical site and bond percolation models describe connectivity by randomly occupying predefined sites or bonds, whereas continuum percolation considers randomly distributed geometrical objects such as discs, spheres, or rods \cite{essam1980percolation}. These approaches provide powerful descriptions of connectivity transitions, but the occupation probability or particle size is generally prescribed independently of the physical process responsible for material formation. Growth-percolation models provide a more direct connection to gelation by allowing initially separated domains to grow until contacts form \cite{herrmann1986geometrical, dodds2002packing, tsakiris2010percolation}. In many existing formulations, however, the growth kinetics and final particle size are introduced independently. This separates the geometrical evolution of the system from the finite amount of reactive material available to each growing domain.

Recent studies increasingly describe gelation as a kinetic, non-equilibrium connectivity transition rather than as a purely static percolation problem. Rouwhorst et al. \cite{Rouwhorst2020} demonstrated that attractive-colloid gelation can be interpreted as a non-equilibrium percolation transition governed by particle-scale association and dissociation kinetics. Aggregation-based descriptions have further shown that the appearance of a system-spanning cluster does not necessarily coincide with completion of network formation. Cook et al. \cite{Cook2023} distinguished the onset of percolation from subsequent incorporation and restructuring within protein hydrogel networks and demonstrated a strong dependence of the gelation kinetics on diffusion- and reaction-limited aggregation. More recently, Haghighi et al. \cite{Haghighi2025} identified percolation, coarsening, and network maturation as distinct stages in the evolution of
attractive colloidal gels. Kinetic effects can also produce departures from predictions based on classical random percolation, as demonstrated for end-linked polymer networks \cite{Beech2023}. These approaches, however, predominantly describe the evolution of connectivity through particle or cluster aggregation, bond formation, or structural rearrangement. A description in which the domains themselves grow by consuming a finite local material reservoir, such that mass conservation determines both the transient growth driving force and the limiting domain size while contact independently governs network topology, remains largely unexplored.

In this letter, polymer gelation is formulated as a locally mass-conserving growth-percolation process. Each nucleus is associated with a finite material reservoir defined by its Voronoi capture zone. Monomer consumption within this reservoir progressively reduces the driving force for growth and determines the limiting particle size without requiring an independently prescribed saturation radius. The same local conservation principle is coupled to interface-controlled and diffusion-controlled \cite{ratke2002growth}, as well as polymer-blob-controlled growth \cite{degennes1979scaling, rubinstein2003polymer}, while contacts between growing particles define an evolving network whose first system-spanning cluster identifies the gel point. The formulation therefore connects nucleation, local material availability, growth kinetics, and global connectivity within a single physical framework.

Consider nuclei at positions $\bm{x}_i$ within a domain $\Omega\subset\mathbb{R}^{d}$, where $d\in\{2,3\}$. The domain is partitioned into bounded Voronoi regions $\mathcal{V}_i$, each of which defines the local material reservoir associated with nucleus $i$. Because the Voronoi regions are generally irregular polygons or polyhedra, each is represented for the purpose of the local growth law by an equivalent $d$-dimensional ball having the same measure. The corresponding equivalent capture radius is

\begin{equation}
R_{e,i}=\left(\frac{V_{V,i}}{c_d}\right)^{1/d},
\label{eq:equivalent_radius}
\end{equation}

\noindent where $V_{V,i}=|\mathcal{V}_i|$, $c_2=\pi$, and $c_3=4\pi/3$. The actual Voronoi geometry therefore determines the amount of locally available material, while the equivalent circular or spherical geometry provides a tractable representation of the corresponding growth kinetics. Introducing the normalized particle radius

\begin{equation}
\xi_i(t)=\frac{R_i(t)}{R_{e,i}},
\label{eq:normalized_radius}
\end{equation}

\noindent the polymer-rich phase occupies a fraction $\xi_i^d$ of the equivalent capture zone, while the remaining solvent occupies the fraction $1-\xi_i^d$. Let $c_{0,i}$ denote the initial monomer concentration assigned to capture zone $i$, $c_p$ the effective monomer concentration in the polymer-rich phase, and $\bar{c}_i(t)$ the spatially averaged concentration in the remaining solvent. Conservation of monomer within each local reservoir gives

\begin{equation}
c_{0,i}=c_p\xi_i^d(t)+\bar{c}_i(t)\left[1-\xi_i^d(t)\right].
\label{eq:local_mass_balance}
\end{equation}

\noindent Defining the initial dimensionless supersaturation as

\begin{equation}
S_{0,i}=\frac{c_{0,i}-c_e}{c_p-c_e},
\label{eq:initial_supersaturation}
\end{equation}

\noindent where $c_e$ is the equilibrium concentration in the solvent, Eq.~\eqref{eq:local_mass_balance} leads directly to the instantaneous local supersaturation

\begin{equation}
\mathcal{S}_i(t)=\frac{S_{0,i}-\xi_i^d(t)}{1-\xi_i^d(t)},
\label{eq:local_supersaturation}
\end{equation}

\noindent which provides the common driving force for all growth mechanisms considered here. During the initial time period, when the particle occupies only a small fraction of its capture zone, $\mathcal{S}_i\simeq S_{0,i}$ and the local reservoir behaves approximately as an undepleted source. As the monomer is incorporated into the polymer-rich phase, however, the available concentration decreases continuously and the driving force for further growth vanishes as local equilibrium is approached. The limiting particle size follows directly from this depletion condition. Growth ceases when $\mathcal{S}_i=0$, giving

\begin{equation}
R_{i,\max}=R_{e,i}S_{0,i}^{1/d}=\left(\frac{S_{0,i}V_{V,i}}{
c_d}\right)^{1/d}.
\label{eq:limiting_radius}
\end{equation}

\noindent The final particle measure therefore satisfies

\begin{equation}
V_{p,i}^{\max}=S_{0,i}V_{V,i}.
\label{eq:limiting_particle_measure}
\end{equation}

\noindent Thus, the heterogeneity of the final particle-size distribution arises directly from the distribution of Voronoi capture-zone measures rather than from an imposed distribution of saturation radii. For a spatially homogeneous precursor solution with $S_{0,i}=S_0$, summation over the complete Voronoi partition yields

\begin{equation}
\phi_{p,\infty}=S_0,
\label{eq:limiting_polymer_fraction}
\end{equation}

\noindent where $\phi_{p,\infty}$ is the asymptotic mass-based polymer fraction. The final amount of polymer is therefore fixed by the initial supersaturation, whereas its spatial distribution is controlled by the nucleation geometry. The particular kinetic mechanism determines how the system approaches this common mass-conserving limit. For interface-controlled growth, monomer transport is sufficiently rapid that attachment or reaction at the particle-solvent interface limits the growth rate. The normalized evolution equation is

\begin{equation}
\frac{\mathrm{d}\xi_i}{\mathrm{d}\tau_i^{\mathrm{int}}}
=\frac{S_{0,i}-\xi_i^d}{1-\xi_i^d},
\label{eq:interface_growth}
\end{equation}

\noindent with $\tau_i^{\mathrm{int}}=k_it/R_{e,i}$. During the initial time period, this gives $R_i-R_{i,0}\propto t$. When transport through the surrounding solvent is rate limiting, the diffusion-controlled growth law for the 3-d equivalent-sphere construction becomes

\begin{equation}
\frac{\mathrm{d}\xi_i}{\mathrm{d}\tau_i^{\mathrm{diff}}}
=\frac{S_{0,i}-\xi_i^3}{\xi_i(1-\xi_i)^2(\xi_i+2)},
\label{eq:diffusion_growth}
\end{equation}

\noindent where $\tau_i^{\mathrm{diff}}=2D_it/R_{e,i}^2$, yielding the initial-time-period behavior $R_i^2-R_{i,0}^2\propto t$. Finally, for compact polymer-blob-controlled growth,

\begin{equation}
\frac{\mathrm{d}\xi_i}{\mathrm{d}\tau_i^{\mathrm{blob}}}
=\frac{1}{3\xi_i^2}\left[\frac{S_{0,i}-\xi_i^3}{1-\xi_i^3}
\right]^3,
\label{eq:blob_growth}
\end{equation}

\noindent with $\tau_i^{\mathrm{blob}}=k_{f,i}^3t/R_{e,i}^3$, leading to $R_i^3-R_{i,0}^3\propto t$ at early times. The three limiting mechanisms may therefore be summarized by \cite{ratke2002growth, degennes1979scaling, rubinstein2003polymer}

\begin{equation}
R_i^q-R_{i,0}^q\propto t,\qquad q=
\begin{cases}
1, & \text{interface-controlled},\\
2, & \text{diffusion-controlled},\\
3, & \text{polymer-blob-controlled}.
\end{cases}
\label{eq:growth_scaling}
\end{equation}

These power laws characterize only the initial growth regime. During the later part of the time period, depletion of the finite local reservoir causes all three mechanisms to deviate from their unconstrained scaling behavior and approach the same limiting radius given by Eq.~\eqref{eq:limiting_radius}. The kinetic mechanism therefore controls the temporal pathway of microstructural evolution, whereas local mass conservation determines the accessible final particle sizes. As the particles grow, contacts are introduced according to the geometrical criterion

\begin{equation}
d_{ij}\leq\lambda_c\left[R_i(t)+R_j(t)\right],
\label{eq:contact_criterion}
\end{equation}

\noindent where $d_{ij}=|\mathbf{x}_i-\mathbf{x}_j|$ is the distance between nuclei $i$ and $j$, and $\lambda_c$ is an effective contact factor. For $\lambda_c=1$, Eq.~\eqref{eq:contact_criterion} corresponds to geometrical touching or overlap, while values slightly larger than unity may represent diffuse interfaces, neck formation, or a finite interaction range. Importantly, $\lambda_c$ modifies connectivity alone and does not affect the local mass balance or limiting particle size.

The resulting contacts define an evolving undirected graph whose nodes are the active nuclei and whose edges correspond to particle contacts. Because the particle centers remain fixed and their radii increase monotonically, the contact network evolves through the progressive addition of edges. Gelation is identified with the first appearance of a connected component spanning the prescribed boundaries of the computational domain. Denoting the corresponding spanning indicator by $\mathcal{P}(t)$, the gelation time is

\begin{equation}
t_g=\inf\left\{t\geq0:\mathcal{P}(t)=1\right\},
\label{eq:gelation_time}
\end{equation}

\noindent which separates gelation from local saturation. The formation of a spanning network generally occurs while at least part of the available monomer remains in the surrounding solvent, such that

\begin{equation}
R_i(t_g)<R_{i,\max}
\label{eq:gel_before_saturation}
\end{equation}

\noindent for at least a subset of the particles. Particle growth and contact formation therefore continue after the gel point. The interval between the first spanning event and local saturation is interpreted as post-gel aging, during which the network becomes progressively more connected and geometrically mature without changing the existence of the system-spanning cluster. A particularly useful measure of this separation is the fraction of the final polymer mass already formed at the gel point. For a spatially homogeneous initial supersaturation, for which $\phi_{p,\infty}=S_0$, this quantity is

\begin{equation}
\chi_{M,g}=\frac{\phi_p(t_g)}{\phi_{p,\infty}}=\frac{\phi_p(t_g)}{S_0}.
\label{eq:mass_fraction_at_gel}
\end{equation}

The quantity $\chi_{M,g}$ therefore measures the degree of material conversion required to establish macroscopic connectivity, whereas

\begin{equation}
1-\chi_{M,g}
\label{eq:post_gel_mass_fraction}
\end{equation}

\noindent represents the fraction of the ultimately available polymer incorporated after the gel point. Gelation and saturation consequently constitute distinct physical events: The former is a topological transition associated with the first system-spanning cluster, while the latter corresponds to exhaustion of the local growth reservoirs. Post-gel aging occupies the regime between these limits, where continued material incorporation generates additional contacts and further develops the topology of an already connected network. This distinction also separates the amount of polymer present from the topology generated by that material. The mass-based polymer fraction is

\begin{equation}
\phi_p(t)=\frac{c_d\sum_i R_i^d(t)}{|\Omega|},
\label{eq:polymer_fraction}
\end{equation}

\noindent whereas the geometrical union fraction counts every point occupied by at least one particle only once. Consequently, overlapping particle domains contribute independently to the mass balance but only once to the geometrically occupied volume. This distinction allows the locally conserved material description to remain consistent while retaining a direct geometrical measure that can be compared with segmented microscopy or tomography.

The framework therefore describes gel formation through two coupled but physically distinct processes. Local conservation determines the depletion of each finite material reservoir and hence the evolution and limiting size of the growing polymer-rich domains, whereas percolation of the corresponding contact network determines the emergence of macroscopic connectivity. Different growth mechanisms modify the time required to reach the connectivity transition but remain constrained by the same locally available material. Gelation consequently represents neither complete conversion nor complete particle growth, but the first topological transition within an evolving, mass-conserving system. This provides a direct physical connection between nucleation conditions, local precursor availability, growth kinetics, gelation, and subsequent network aging.

\begin{figure*}[ht!]
    \centering
    \includegraphics[width=\linewidth]{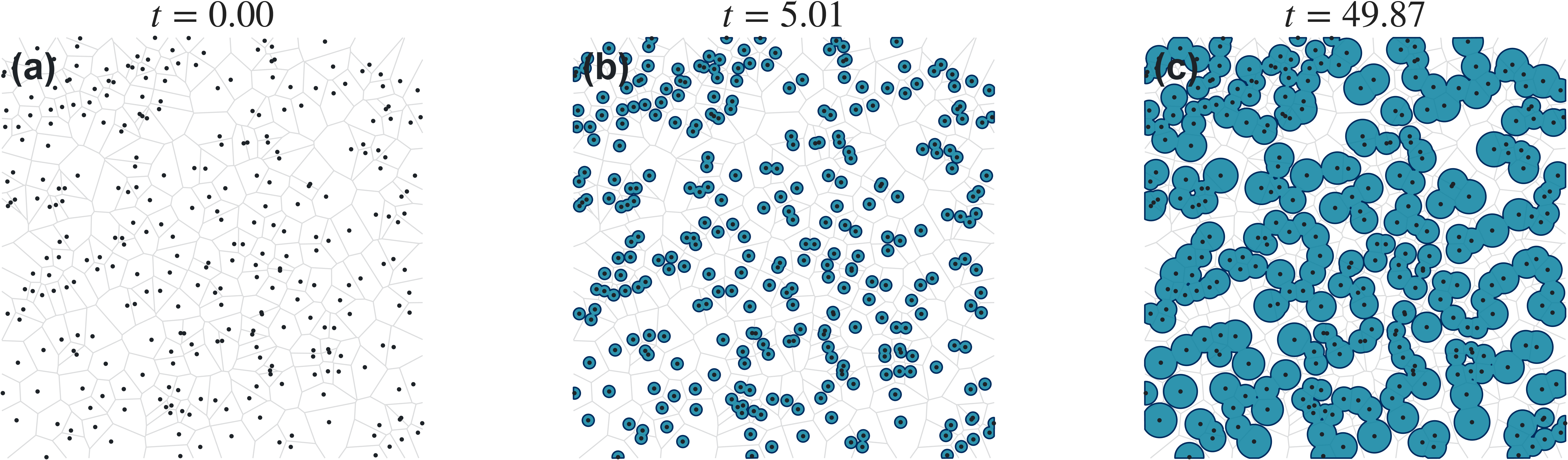}\\[2mm]
    \includegraphics[width=\linewidth]{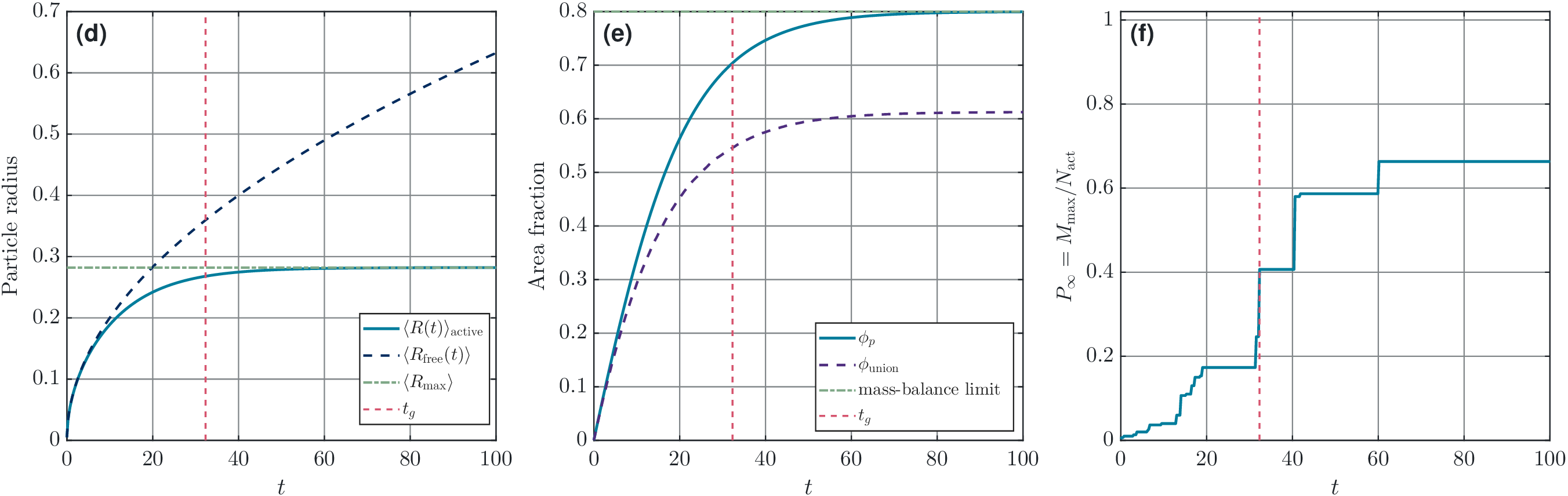}
    \caption{Evolution of the locally mass-conserving growth-percolation model in 2-d for a representative realization using diffusion-controlled growth. (a-c) Evolution of the initially isolated nuclei into an interconnected polymer-rich structure. The grey lines indicate the Voronoi capture zones, which define the finite local material reservoirs.
(d) Mean particle radius compared with unconstrained growth and the mean
mass-conserving limiting radius. (e) Mass-based polymer fraction $\phi_p$ and geometrical union fraction $\phi_{\mathrm{union}}$; the former approaches the mass-balance limit prescribed by the initial supersaturation.
(f) Largest-cluster fraction $P_\infty$ as the growing domains progressively
form a connected network. The indicated $t_g$ denotes the first system-spanning state. Particle growth continues beyond $t_g$, illustrating
the separation between gelation and local saturation.
}
    \label{fig:2dgrowth}
\end{figure*}

The evolution predicted by the proposed growth-percolation framework is first illustrated in Fig.~1 for a representative 2-d realization using diffusion-controlled growth. An initially disconnected population of nuclei evolves through local particle growth and progressive contact formation into a system-spanning network. The snapshots emphasize that the connectivity transition is not imposed independently of the growth process, but emerges naturally from the competition between the spatial arrangement of the nuclei, the finite amount of material available within their Voronoi capture zones, and the resulting time-dependent particle radii. The same formulation therefore simultaneously describes the evolution of the characteristic structural length scale and the topology of the emerging network.

The 2-d example in Fig.~1(a-c) demonstrates particularly clearly the role of local mass conservation in limiting particle growth. Starting from an initially dilute distribution of nuclei, the particles grow rapidly and progressively establish contacts with their neighbors. At early times, as observed in Fig.~1(d), the evolution closely follows the corresponding unconstrained growth behavior because only a small fraction of the local material reservoirs has been consumed. As growth proceeds, however, depletion within the individual Voronoi capture zones reduces the local supersaturation and causes the actual mean particle radius to depart increasingly from the unconstrained growth trajectory. Whereas the latter continues to increase, the mass-conserving solution approaches the mean thermodynamic saturation radius. The limiting radius is therefore not imposed as an additional empirical parameter, but emerges directly from the initial supersaturation and the amount of material assigned to each nucleus through its Voronoi capture zone.

The same behavior is reflected at the level of the total polymer fraction. Figure~1(e) demonstrates that the mass-based polymer fraction increases monotonically and approaches the thermodynamic mass-balance limit prescribed by the initial supersaturation. This provides a direct numerical manifestation of the analytical result $\phi_{p,\infty}=S_0$ for a spatially homogeneous precursor solution. In contrast, the geometrical union fraction remains systematically below the mass-based polymer fraction once appreciable particle overlap develops. The separation between these quantities is a direct consequence of their different physical meanings: $\phi_p$ accounts for the polymer mass associated with each local reservoir independently, whereas $\phi_{\mathrm{union}}$ counts an overlapping spatial region only once. Their divergence therefore provides a geometrical measure of the increasing overlap and interpenetration of the growing domains while preserving the underlying local mass balance.

Importantly, the formation of a gel occurs substantially before the system reaches its final mass-conserving state. In Fig.~1(f), the first spanning cluster appears at the indicated gelation time $t_g$, while both particle growth and the polymer fraction continue to evolve thereafter. The largest-cluster fraction exhibits a rapid increase around this point, demonstrating the transition from a collection of finite clusters to a macroscopically connected structure. Gelation is therefore not associated with exhaustion of the available material, but with the first topological event at which a connected path extends across the system. This distinction is central to the present description: The growth dynamics determine when sufficient contacts have formed, whereas the gel point itself is identified solely through connectivity.
The subsequent evolution illustrates the post-gel aging regime contained naturally within the model. After $t_g$, further incorporation of monomer increases the particle radii and generates additional contacts within a network that is already spanning. Consequently, the gel point and the final saturation state represent two distinct stages of structural development. For a homogeneous initial supersaturation, the fraction of the ultimately available polymer already incorporated at gelation can be quantified through Eq. \ref{eq:mass_fraction_at_gel}. Accordingly, $1-\chi_{M,g}$ corresponds to the fraction of the final polymer mass incorporated during post-gel aging. The simulations therefore provide a direct physical interpretation of aging as continued material incorporation and topological maturation following the initial percolation transition, rather than as a separate phenomenological process imposed after gelation.

\begin{figure*}
    \centering
    \includegraphics[width=\linewidth]{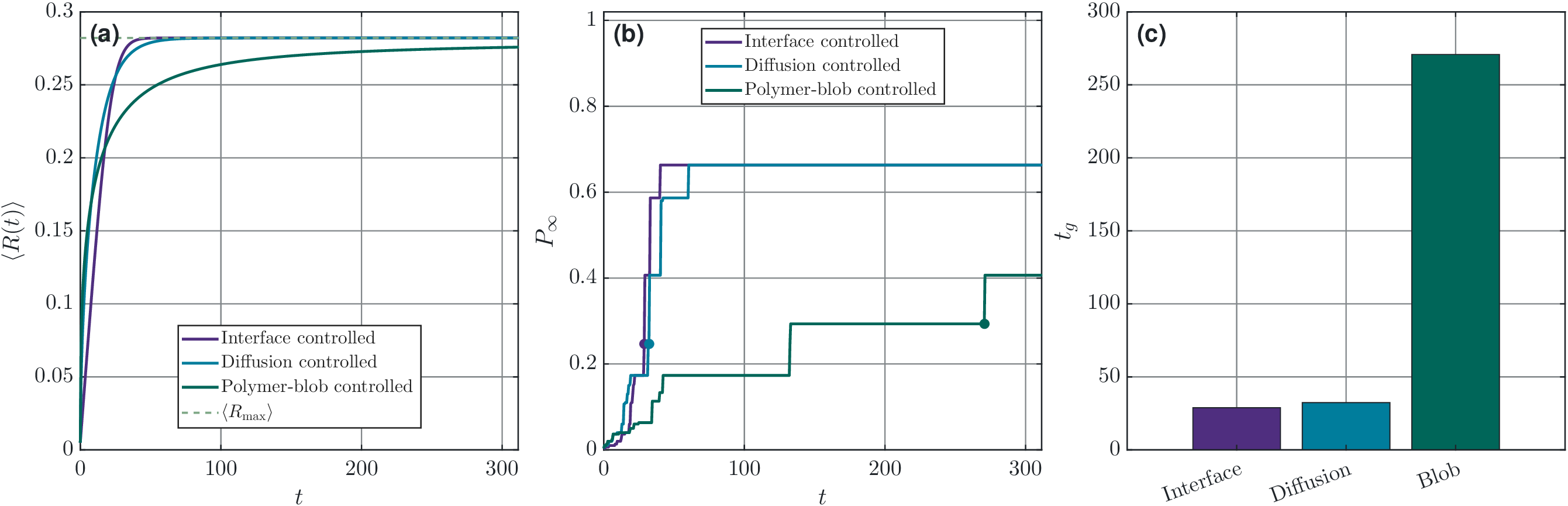}\\[2mm]
    \includegraphics[width=\linewidth]{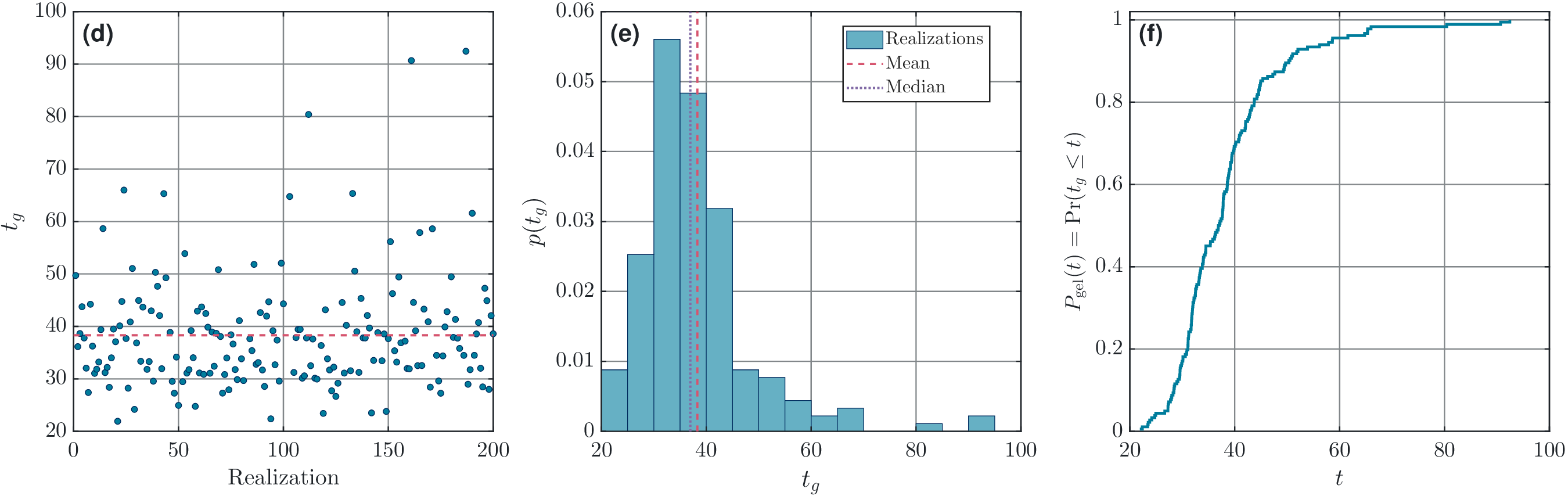}
    \caption{
Influence of growth kinetics and stochastic nucleation geometry on the
gelation time in 2-d. (a-c) Comparison of interface-controlled, diffusion-controlled, and polymer-blob-controlled growth applied to exactly the same realization of growth sites. The nucleus positions, Voronoi capture zones, initial supersaturation, limiting particle radii, and contact criterion are
identical in all three cases; only the kinetic law and its corresponding
kinetic time scale are changed. (a) Evolution of the mean particle radius.
(b) Largest-cluster fraction $P_\infty$, with symbols indicating the first
system-spanning state. (c) Corresponding gelation times. For the representative kinetic parameters considered here, gelation occurs at approximately $t_g=28.86$, $32.33$, and $270.79$ for interface-controlled,
diffusion-controlled, and polymer-blob-controlled growth, respectively.
(d-f) Realization-to-realization variation obtained by keeping the number
of growth sites and all material and kinetic parameters fixed while
resampling only their spatial positions. (d) Gelation time for the individual realizations. (e) Distribution of gelation times $p(t_g)$.
(f) Empirical cumulative gelation probability $P_{\mathrm{gel}}(t)=\Pr(t_g\le t)$.
}
    \label{fig:2danalysis}
\end{figure*}

The role of the rate-controlling growth mechanism can be isolated by applying the three kinetic laws to exactly the same realization of nuclei, as shown in Figs.~2(a-c). In this comparison, the nucleus positions and hence the Voronoi capture zones, local material reservoirs, limiting particle radii, and all geometrical contact conditions remain unchanged. The different gelation times therefore originate from the temporal growth kinetics rather than from differences in the accessible network geometry. For the representative kinetic scales considered here, interface-controlled growth produces the earliest spanning network at $t_g=28.86$, followed by diffusion-controlled growth at $t_g=32.33$, whereas polymer-blob-controlled growth delays gelation until $t_g=270.79$. The strong retardation originates from the cubic dependence of the blob-growth rate on the local supersaturation. As the finite local reservoirs become depleted,
$\mathcal{S}_i\rightarrow0$, the blob growth rate decreases as $\mathcal{S}_i^3$, whereas the interface- and diffusion-controlled rates
remain linearly proportional to $\mathcal{S}_i$. The distinction between
the mechanisms therefore becomes particularly pronounced during the later
stages of growth approaching the percolation transition. The rate-controlling mechanism can therefore alter the time required to establish macroscopic connectivity by around a factor of ten even for an identical initial spatial configuration. Despite this pronounced separation of the gelation times, the three systems remain constrained by the same locally available material and therefore approach the same particle-specific limiting radii. This directly demonstrates the distinction anticipated by Eqs.~\eqref{eq:interface_growth}-\eqref{eq:growth_scaling}: the kinetic mechanism determines the temporal pathway toward network formation, whereas local mass conservation determines the accessible final state. The absolute ordering of dimensional gelation times additionally depends on the material-specific kinetic coefficients; the present comparison
isolates the consequence of the different growth-law forms after normalization to a common reference time scale.

A second source of variability arises from the stochastic spatial
distribution of the nuclei. Figures~2(d-f) isolate this contribution by holding the number of growth sites,  supersaturation, kinetic parameters, and contact criterion fixed while independently resampling only the nucleus positions. Each realization in Fig.~2(d) consequently represents the same nominal macroscopic synthesis conditions but generates a different Voronoi partition and contact geometry. The gelation time therefore becomes a
realization-dependent quantity rather than a single deterministic value. Spatial fluctuations influence gelation in two coupled ways. First, the Voronoi tessellation redistributes the finite local material reservoirs and therefore changes the particle-specific limiting radii. Second, variation in the distances and arrangement of neighboring nuclei changes the geometrical pathways through which a system-spanning cluster can develop.
The resulting distribution $p(t_g)$ in Fig.~2(e) therefore represents a purely microstructural contribution to the variability of the gel point. For the present ensemble, the mean gelation time is $\langle t_g\rangle=38.31$, with a standard deviation $\sigma_{t_g}=10.77$ and coefficient of variation $\sigma_{t_g}/\langle t_g\rangle=0.28$.

This stochasticity provides a possible microstructural contribution to the experimentally observed spread in gelation times between nominally identical samples. Even samples taken from the same precursor batch may not develop identical microscopic nucleation patterns and may consequently reach the
first system-spanning state at different times. Additional experimental variability associated with mixing, temperature, catalyst distribution, or composition would further broaden this distribution. Exemplarily, experimental measurements of base-catalyzed resorcinol-formaldehyde gels have shown appreciable sample-to-sample scatter in gelation time under nominally identical conditions. Ratke and Hadjuk \cite{ratke2015size} distinguished the onset of gelation, identified by the transition from a clear to a turbid droplet,
from a later final gelation time associated with complete opacity, and reported measurable variations in both quantities for repeated experiments at fixed droplet size.

\begin{figure*}
    \centering
    \includegraphics[width=\linewidth]{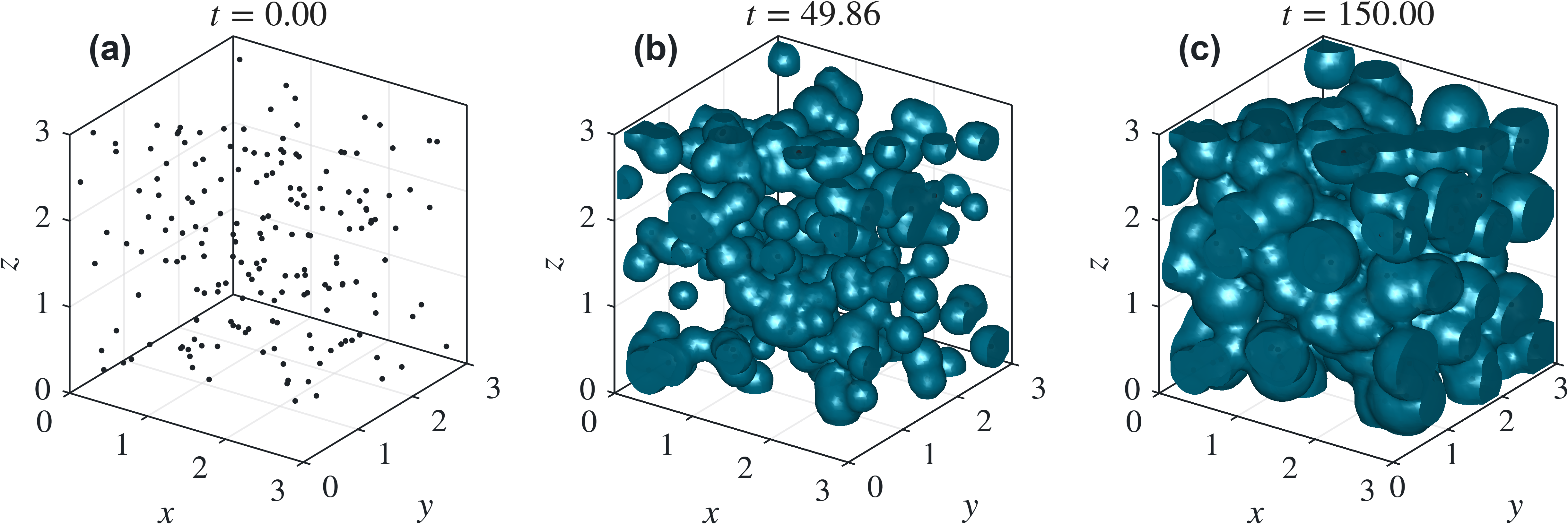}\\[2mm]
    \includegraphics[width=\linewidth]{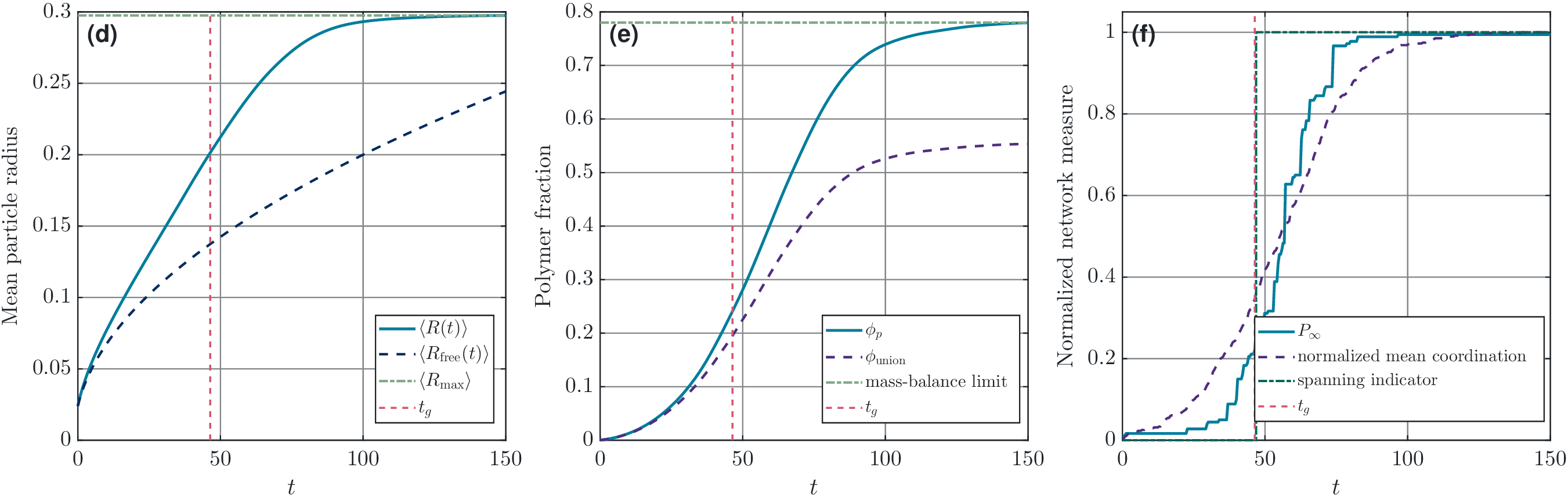}
    \caption{
Growth-percolation evolution in 3-d for a representative realization using
diffusion-controlled growth. (a-c) Development of initially isolated spherical domains into an interconnected three-dimensional network.
(d) Mean particle radius and the corresponding mass-conserving limiting
radius. (e) Mass-based and geometrical polymer fractions. (f) Evolution of the network connectivity, showing that the first spanning state precedes completion of particle growth and network maturation.}
    \label{fig:placeholder}
\end{figure*}

The representative 3-d simulation in Fig.~3 demonstrates that the same
mass-conserving growth-percolation framework carries directly to a 3-d geometry. The initially isolated nuclei grow into approximately spherical polymer-rich domains and progressively generate an interconnected 3-d structure (see Fig.~3(a-c)). The mean particle radius remains below the thermodynamic saturation radius over the time interval shown, as can be observed from Fig.~3d, demonstrating that the first system-spanning network forms while substantial growth potential is still available. Consistently, the mass-based polymer fraction at the gel point remains well below its thermodynamic limiting value, as illustrated in Fig.~3e. Thus, only a fraction of the ultimately available polymer is required to establish global connectivity in this realization.

The difference between the mass-based and geometrical polymer fractions also becomes increasingly pronounced during the 3-d evolution. As neighboring particles overlap, the independently conserved material associated with the local reservoirs continues to increase the mass-based fraction, whereas overlapping regions contribute only once to the geometrical union. This distinction is particularly relevant for relating the model to experimentally reconstructed microstructures. The mass-based fraction represents the conserved amount of polymer predicted from the precursor composition, while the union fraction corresponds more directly to the solid fraction that would be obtained from a segmented 2- or 3-d image.

The network formation graph in Fig.~3f further demonstrates that the first spanning event does not represent completion of the connectivity evolution. The spanning indicator changes abruptly at $t_g$, as required by the geometrical definition of gelation, whereas both the largest-cluster fraction and the normalized mean coordination continue to increase beyond this point. Immediately after gelation, the spanning component therefore contains only part of the final network. Continued growth progressively incorporates additional particles into the dominant cluster and increases the number of contacts per particle. The model consequently distinguishes naturally between the onset of global connectivity and the subsequent maturation of that connectivity. The latter is particularly significant in 3-d, where a system may acquire a spanning path while remaining far from its final coordination state.

In summary, Figs.~1-3 show that the amount of growth required for gelation is not a universal fraction of the final state. The gel point depends on the kinetic pathway as well as on the particular spatial
realization of the nuclei and its associated capture-zone topology.
A favorable sequence of contacts may establish a spanning path while a
substantial fraction of the local reservoirs remains unconsumed, whereas
another realization or a slower kinetic pathway may require considerably
longer evolution before percolation occurs. A second distinctive feature is that transient growth and final particle size are governed by the same physical constraint. In many growth-percolation descriptions, the kinetic growth law and saturation radius constitute separate ingredients. Here, the decreasing local supersaturation simultaneously drives the transient evolution and arrests growth at the mass-conserving limiting radius. In the present formulation, the decreasing supersaturation (Eq. \ref{eq:local_supersaturation})
simultaneously drives the transient growth and causes that growth to vanish according to Eq. \ref{eq:limiting_radius}.
The final microstructural length scale is therefore determined by the same local material conservation that governs its temporal evolution. Moreover, because $R_{e,i}$ follows directly from the Voronoi capture-zone measure, spatial heterogeneity in the nucleation pattern is translated naturally into heterogeneity of the final particle sizes without prescribing an additional particle-size distribution.  Figure~2(a-c) illustrates the consequence directly; markedly different kinetic pathways and gelation times occur while the underlying limiting particle sizes remain unchanged.

In this sense, the present framework does not prescribe a critical particle size or polymer fraction for gelation. Instead, the critical state emerges from the evolving geometry and topology of each realization. This provides an important distinction from conventional continuum-percolation descriptions. In classical models, particle size, occupation probability, or solid fraction typically acts as an externally varied control parameter, and percolation is detected as that parameter crosses a critical value \cite{lorenz2001precise, consiglio2003continuum}. Here, the geometrical control parameter itself evolves according to a local physical growth law. Particle size is therefore not independently prescribed: It follows from monomer consumption within a finite Voronoi-associated reservoir, while the same mass balance determines the maximum radius accessible to every particle. Percolation is consequently embedded within the material-formation process rather than applied to a separately generated geometry.

Most importantly, the framework establishes a direct connection between three physical levels that are usually treated separately: Local material availability, mesoscale domain growth, and global network topology. The precursor composition determines the available polymer mass; the Voronoi tessellation distributes this finite material locally among the nuclei; the selected kinetic mechanism determines how rapidly the corresponding domains grow; and the evolving contact graph determines when macroscopic connectivity first appears. The results further show that the gelation time is controlled independently by both the kinetic pathway and the stochastic nucleation geometry, while the accessible final state remains constrained by local mass conservation. Gelation and ageing then arise as two distinct stages of the same continuous evolution: gelation is the first topological transition to a spanning network, whereas aging describes continued mass incorporation and contact formation as the system approaches local saturation. This coupling of finite-reservoir growth, local mass conservation, and dynamic percolation constitutes the central novelty of the model and provides a physically interpretable route from synthesis conditions to particle size, gelation time, network connectivity, and post-gel structural evolution. Such a description can naturally be coupled with cluster-cluster aggregation or diffusion-limited aggregation models \cite{witten1981diffusion, meakin1983formation} to additionally account for translational/rotational motion, collision, and restructuring of particles and clusters. This would extend the present growth-dominated framework toward systems in which particle growth and aggregate transport act concurrently during gelation.

\nocite{*}

\bibliography{apssamp}% Produces the bibliography via BibTeX.

\begin{thebibliography}{17}%
\makeatletter
\providecommand \@ifxundefined [1]{%
 \@ifx{#1\undefined}
}%
\providecommand \@ifnum [1]{%
 \ifnum #1\expandafter \@firstoftwo
 \else \expandafter \@secondoftwo
 \fi
}%
\providecommand \@ifx [1]{%
 \ifx #1\expandafter \@firstoftwo
 \else \expandafter \@secondoftwo
 \fi
}%
\providecommand \natexlab [1]{#1}%
\providecommand \enquote  [1]{``#1''}%
\providecommand \bibnamefont  [1]{#1}%
\providecommand \bibfnamefont [1]{#1}%
\providecommand \citenamefont [1]{#1}%
\providecommand \href@noop [0]{\@secondoftwo}%
\providecommand \href [0]{\begingroup \@sanitize@url \@href}%
\providecommand \@href[1]{\@@startlink{#1}\@@href}%
\providecommand \@@href[1]{\endgroup#1\@@endlink}%
\providecommand \@sanitize@url [0]{\catcode `\\12\catcode `\$12\catcode
  `\&12\catcode `\#12\catcode `\^12\catcode `\_12\catcode `\%12\relax}%
\providecommand \@@startlink[1]{}%
\providecommand \@@endlink[0]{}%
\providecommand \url  [0]{\begingroup\@sanitize@url \@url }%
\providecommand \@url [1]{\endgroup\@href {#1}{\urlprefix }}%
\providecommand \urlprefix  [0]{URL }%
\providecommand \Eprint [0]{\href }%
\providecommand \doibase [0]{https://doi.org/}%
\providecommand \selectlanguage [0]{\@gobble}%
\providecommand \bibinfo  [0]{\@secondoftwo}%
\providecommand \bibfield  [0]{\@secondoftwo}%
\providecommand \translation [1]{[#1]}%
\providecommand \BibitemOpen [0]{}%
\providecommand \bibitemStop [0]{}%
\providecommand \bibitemNoStop [0]{.\EOS\space}%
\providecommand \EOS [0]{\spacefactor3000\relax}%
\providecommand \BibitemShut  [1]{\csname bibitem#1\endcsname}%
\let\auto@bib@innerbib\@empty
%</preamble>
\bibitem [{\citenamefont {Ratke}\ and\ \citenamefont
  {Gurikov}(2021)}]{ratke2021chemistry}%
  \BibitemOpen
  \bibfield  {author} {\bibinfo {author} {\bibfnamefont {L.}~\bibnamefont
  {Ratke}}\ and\ \bibinfo {author} {\bibfnamefont {P.}~\bibnamefont
  {Gurikov}},\ }\href@noop {} {\emph {\bibinfo {title} {The chemistry and
  physics of aerogels: synthesis, processing, and properties}}}\ (\bibinfo
  {publisher} {Cambridge University Press},\ \bibinfo {year}
  {2021})\BibitemShut {NoStop}%
\bibitem [{\citenamefont {Essam}(1980)}]{essam1980percolation}%
  \BibitemOpen
  \bibfield  {author} {\bibinfo {author} {\bibfnamefont {J.~W.}\ \bibnamefont
  {Essam}},\ }\bibfield  {title} {\bibinfo {title} {Percolation theory},\
  }\href@noop {} {\bibfield  {journal} {\bibinfo  {journal} {Reports on
  Progress in Physics}\ }\textbf {\bibinfo {volume} {43}},\ \bibinfo {pages}
  {833} (\bibinfo {year} {1980})}\BibitemShut {NoStop}%
\bibitem [{\citenamefont {Herrmann}(1986)}]{herrmann1986geometrical}%
  \BibitemOpen
  \bibfield  {author} {\bibinfo {author} {\bibfnamefont {H.~J.}\ \bibnamefont
  {Herrmann}},\ }\bibfield  {title} {\bibinfo {title} {Geometrical cluster
  growth models and kinetic gelation},\ }\href@noop {} {\bibfield  {journal}
  {\bibinfo  {journal} {Physics Reports}\ }\textbf {\bibinfo {volume} {136}},\
  \bibinfo {pages} {153} (\bibinfo {year} {1986})}\BibitemShut {NoStop}%
\bibitem [{\citenamefont {Dodds}\ and\ \citenamefont
  {Weitz}(2002)}]{dodds2002packing}%
  \BibitemOpen
  \bibfield  {author} {\bibinfo {author} {\bibfnamefont {P.~S.}\ \bibnamefont
  {Dodds}}\ and\ \bibinfo {author} {\bibfnamefont {J.~S.}\ \bibnamefont
  {Weitz}},\ }\bibfield  {title} {\bibinfo {title} {Packing-limited growth},\
  }\href@noop {} {\bibfield  {journal} {\bibinfo  {journal} {Physical Review
  E}\ }\textbf {\bibinfo {volume} {65}},\ \bibinfo {pages} {056108} (\bibinfo
  {year} {2002})}\BibitemShut {NoStop}%
\bibitem [{\citenamefont {Tsakiris}\ \emph {et~al.}(2010)\citenamefont
  {Tsakiris}, \citenamefont {Maragakis}, \citenamefont {Kosmidis},\ and\
  \citenamefont {Argyrakis}}]{tsakiris2010percolation}%
  \BibitemOpen
  \bibfield  {author} {\bibinfo {author} {\bibfnamefont {N.}~\bibnamefont
  {Tsakiris}}, \bibinfo {author} {\bibfnamefont {M.}~\bibnamefont {Maragakis}},
  \bibinfo {author} {\bibfnamefont {K.}~\bibnamefont {Kosmidis}},\ and\
  \bibinfo {author} {\bibfnamefont {P.}~\bibnamefont {Argyrakis}},\ }\bibfield
  {title} {\bibinfo {title} {Percolation of randomly distributed growing
  clusters: finite-size scaling and critical exponents for the square
  lattice},\ }\href@noop {} {\bibfield  {journal} {\bibinfo  {journal}
  {Physical Review E}\ }\textbf {\bibinfo {volume} {82}},\ \bibinfo {pages}
  {041108} (\bibinfo {year} {2010})}\BibitemShut {NoStop}%
\bibitem [{\citenamefont {Rouwhorst}\ \emph {et~al.}(2020)\citenamefont
  {Rouwhorst}, \citenamefont {Ness}, \citenamefont {Stoyanov}, \citenamefont
  {Zaccone},\ and\ \citenamefont {Schall}}]{Rouwhorst2020}%
  \BibitemOpen
  \bibfield  {author} {\bibinfo {author} {\bibfnamefont {J.}~\bibnamefont
  {Rouwhorst}}, \bibinfo {author} {\bibfnamefont {C.}~\bibnamefont {Ness}},
  \bibinfo {author} {\bibfnamefont {S.}~\bibnamefont {Stoyanov}}, \bibinfo
  {author} {\bibfnamefont {A.}~\bibnamefont {Zaccone}},\ and\ \bibinfo {author}
  {\bibfnamefont {P.}~\bibnamefont {Schall}},\ }\bibfield  {title} {\bibinfo
  {title} {Nonequilibrium continuous phase transition in colloidal gelation
  with short-range attraction},\ }\href
  {https://doi.org/10.1038/s41467-020-17353-8} {\bibfield  {journal} {\bibinfo
  {journal} {Nature Communications}\ }\textbf {\bibinfo {volume} {11}},\
  \bibinfo {pages} {3558} (\bibinfo {year} {2020})}\BibitemShut {NoStop}%
\bibitem [{\citenamefont {Cook}\ \emph {et~al.}(2023)\citenamefont {Cook},
  \citenamefont {Head},\ and\ \citenamefont {Dougan}}]{Cook2023}%
  \BibitemOpen
  \bibfield  {author} {\bibinfo {author} {\bibfnamefont {K.~R.}\ \bibnamefont
  {Cook}}, \bibinfo {author} {\bibfnamefont {D.}~\bibnamefont {Head}},\ and\
  \bibinfo {author} {\bibfnamefont {L.}~\bibnamefont {Dougan}},\ }\bibfield
  {title} {\bibinfo {title} {Modelling network formation in folded protein
  hydrogels by cluster aggregation kinetics},\ }\href
  {https://doi.org/10.1039/D3SM00111C} {\bibfield  {journal} {\bibinfo
  {journal} {Soft Matter}\ }\textbf {\bibinfo {volume} {19}},\ \bibinfo {pages}
  {2780} (\bibinfo {year} {2023})}\BibitemShut {NoStop}%
\bibitem [{\citenamefont {Haghighi}\ \emph {et~al.}(2025)\citenamefont
  {Haghighi}, \citenamefont {Nabizadeh},\ and\ \citenamefont
  {Jamali}}]{Haghighi2025}%
  \BibitemOpen
  \bibfield  {author} {\bibinfo {author} {\bibfnamefont {P.}~\bibnamefont
  {Haghighi}}, \bibinfo {author} {\bibfnamefont {M.}~\bibnamefont
  {Nabizadeh}},\ and\ \bibinfo {author} {\bibfnamefont {S.}~\bibnamefont
  {Jamali}},\ }\bibfield  {title} {\bibinfo {title} {Emergence and evolution of
  a particulate network during gelation and coarsening of attractive
  colloids},\ }\href {https://doi.org/10.1039/D5SM00572H} {\bibfield  {journal}
  {\bibinfo  {journal} {Soft Matter}\ }\textbf {\bibinfo {volume} {21}},\
  \bibinfo {pages} {7904} (\bibinfo {year} {2025})}\BibitemShut {NoStop}%
\bibitem [{\citenamefont {Beech}\ \emph {et~al.}(2023)\citenamefont {Beech},
  \citenamefont {Lin}, \citenamefont {Sen}, \citenamefont {Rota},\ and\
  \citenamefont {Olsen}}]{Beech2023}%
  \BibitemOpen
  \bibfield  {author} {\bibinfo {author} {\bibfnamefont {H.~K.}\ \bibnamefont
  {Beech}}, \bibinfo {author} {\bibfnamefont {T.-S.}\ \bibnamefont {Lin}},
  \bibinfo {author} {\bibfnamefont {D.}~\bibnamefont {Sen}}, \bibinfo {author}
  {\bibfnamefont {D.}~\bibnamefont {Rota}},\ and\ \bibinfo {author}
  {\bibfnamefont {B.~D.}\ \bibnamefont {Olsen}},\ }\bibfield  {title} {\bibinfo
  {title} {Kinetics of polymer gel formation cause deviation from percolation
  theory in the dilute regime},\ }\href
  {https://doi.org/10.1021/acs.macromol.3c00831} {\bibfield  {journal}
  {\bibinfo  {journal} {Macromolecules}\ }\textbf {\bibinfo {volume} {56}},\
  \bibinfo {pages} {9255} (\bibinfo {year} {2023})}\BibitemShut {NoStop}%
\bibitem [{\citenamefont {Ratke}\ and\ \citenamefont
  {Voorhees}(2002)}]{ratke2002growth}%
  \BibitemOpen
  \bibfield  {author} {\bibinfo {author} {\bibfnamefont {L.}~\bibnamefont
  {Ratke}}\ and\ \bibinfo {author} {\bibfnamefont {P.~W.}\ \bibnamefont
  {Voorhees}},\ }\href@noop {} {\emph {\bibinfo {title} {Growth and coarsening:
  Ostwald ripening in material processing}}}\ (\bibinfo  {publisher} {Springer
  Science \& Business Media},\ \bibinfo {year} {2002})\BibitemShut {NoStop}%
\bibitem [{\citenamefont {de~Gennes}(1979)}]{degennes1979scaling}%
  \BibitemOpen
  \bibfield  {author} {\bibinfo {author} {\bibfnamefont {P.-G.}\ \bibnamefont
  {de~Gennes}},\ }\href@noop {} {\emph {\bibinfo {title} {Scaling concepts in
  polymer physics}}}\ (\bibinfo  {publisher} {Cornell University Press},\
  \bibinfo {address} {Ithaca, NY},\ \bibinfo {year} {1979})\BibitemShut
  {NoStop}%
\bibitem [{\citenamefont {Rubinstein}\ and\ \citenamefont
  {Colby}(2003)}]{rubinstein2003polymer}%
  \BibitemOpen
  \bibfield  {author} {\bibinfo {author} {\bibfnamefont {M.}~\bibnamefont
  {Rubinstein}}\ and\ \bibinfo {author} {\bibfnamefont {R.~H.}\ \bibnamefont
  {Colby}},\ }\href@noop {} {\emph {\bibinfo {title} {Polymer physics}}}\
  (\bibinfo  {publisher} {Oxford University Press, New York},\ \bibinfo {year}
  {2003})\BibitemShut {NoStop}%
\bibitem [{\citenamefont {Ratke}\ and\ \citenamefont
  {Hajduk}(2015)}]{ratke2015size}%
  \BibitemOpen
  \bibfield  {author} {\bibinfo {author} {\bibfnamefont {L.}~\bibnamefont
  {Ratke}}\ and\ \bibinfo {author} {\bibfnamefont {A.}~\bibnamefont {Hajduk}},\
  }\bibfield  {title} {\bibinfo {title} {On the size effect of gelation
  kinetics in rf aerogels},\ }\href@noop {} {\bibfield  {journal} {\bibinfo
  {journal} {Gels}\ }\textbf {\bibinfo {volume} {1}},\ \bibinfo {pages} {276}
  (\bibinfo {year} {2015})}\BibitemShut {NoStop}%
\bibitem [{\citenamefont {Lorenz}\ and\ \citenamefont
  {Ziff}(2001)}]{lorenz2001precise}%
  \BibitemOpen
  \bibfield  {author} {\bibinfo {author} {\bibfnamefont {C.~D.}\ \bibnamefont
  {Lorenz}}\ and\ \bibinfo {author} {\bibfnamefont {R.~M.}\ \bibnamefont
  {Ziff}},\ }\bibfield  {title} {\bibinfo {title} {Precise determination of the
  critical percolation threshold for the three-dimensional “swiss cheese”
  model using a growth algorithm},\ }\href@noop {} {\bibfield  {journal}
  {\bibinfo  {journal} {The Journal of Chemical Physics}\ }\textbf {\bibinfo
  {volume} {114}},\ \bibinfo {pages} {3659} (\bibinfo {year}
  {2001})}\BibitemShut {NoStop}%
\bibitem [{\citenamefont {Consiglio}\ \emph {et~al.}(2003)\citenamefont
  {Consiglio}, \citenamefont {Baker}, \citenamefont {Paul},\ and\ \citenamefont
  {Stanley}}]{consiglio2003continuum}%
  \BibitemOpen
  \bibfield  {author} {\bibinfo {author} {\bibfnamefont {R.}~\bibnamefont
  {Consiglio}}, \bibinfo {author} {\bibfnamefont {D.}~\bibnamefont {Baker}},
  \bibinfo {author} {\bibfnamefont {G.}~\bibnamefont {Paul}},\ and\ \bibinfo
  {author} {\bibfnamefont {H.}~\bibnamefont {Stanley}},\ }\bibfield  {title}
  {\bibinfo {title} {Continuum percolation thresholds for mixtures of spheres
  of different sizes},\ }\href@noop {} {\bibfield  {journal} {\bibinfo
  {journal} {Physica A: Statistical Mechanics and its Applications}\ }\textbf
  {\bibinfo {volume} {319}},\ \bibinfo {pages} {49} (\bibinfo {year}
  {2003})}\BibitemShut {NoStop}%
\bibitem [{\citenamefont {Witten~Jr}\ and\ \citenamefont
  {Sander}(1981)}]{witten1981diffusion}%
  \BibitemOpen
  \bibfield  {author} {\bibinfo {author} {\bibfnamefont {T.~A.}\ \bibnamefont
  {Witten~Jr}}\ and\ \bibinfo {author} {\bibfnamefont {L.~M.}\ \bibnamefont
  {Sander}},\ }\bibfield  {title} {\bibinfo {title} {Diffusion-limited
  aggregation, a kinetic critical phenomenon},\ }\href@noop {} {\bibfield
  {journal} {\bibinfo  {journal} {Physical Review Letters}\ }\textbf {\bibinfo
  {volume} {47}},\ \bibinfo {pages} {1400} (\bibinfo {year}
  {1981})}\BibitemShut {NoStop}%
\bibitem [{\citenamefont {Meakin}(1983)}]{meakin1983formation}%
  \BibitemOpen
  \bibfield  {author} {\bibinfo {author} {\bibfnamefont {P.}~\bibnamefont
  {Meakin}},\ }\bibfield  {title} {\bibinfo {title} {Formation of fractal
  clusters and networks by irreversible diffusion-limited aggregation},\
  }\href@noop {} {\bibfield  {journal} {\bibinfo  {journal} {Physical Review
  Letters}\ }\textbf {\bibinfo {volume} {51}},\ \bibinfo {pages} {1119}
  (\bibinfo {year} {1983})}\BibitemShut {NoStop}%
\end{thebibliography}%

\end{document}